\documentclass[twocolumn]{aastex631}

\usepackage{comment}

\usepackage{CJK,amsmath}

\newcommand{\beq}{\begin{equation}}
\newcommand{\eeq}{\end{equation}}

\shorttitle{Subdwarf B Star Spin}
\shortauthors{Ma \& Fuller}

\graphicspath{{./}{figures/}}

\begin{document}
\begin{CJK*}{UTF8}{gbsn}

\title{Tidal Spin-up of Subdwarf B Stars}
\correspondingauthor{Linhao Ma}
\email{linhaoma@princeton.edu}

\author[0000-0001-6117-5750]{Linhao Ma（马林昊）}
\affiliation{TAPIR, Mailcode 350-17, California Institute of Technology, Pasadena, CA 91125, USA}
\affiliation{Department of Astrophysical Sciences, Princeton University, Princeton, NJ 08544, USA}

\author[0000-0002-4544-0750]{Jim Fuller}
\affiliation{TAPIR, Mailcode 350-17, California Institute of Technology, Pasadena, CA 91125, USA}

\begin{abstract}

Hot subdwarf B (sdB) stars are stripped helium-burning stars that are often found in close binaries, where they experience strong tidal interactions. The dissipation of tidally excited gravity waves alters their rotational evolution throughout the sdB lifetime. While many sdB binaries have well-measured rotational and orbital frequencies, there have been few theoretical efforts to accurately calculate the tidal torque produced by gravity waves. In this work, we directly calculate the tidal excitation of internal gravity waves in realistic sdB stellar models and integrate the coupled spin--orbit evolution of sdB binaries. We find that for canonical sdB ($M_\mathrm{sdB}=0.47\,M_\odot$) binaries, the transitional orbital period below which they could reach tidal synchronization in the sdB lifetime is $\sim \! 0.2\;\mathrm{days}$, with weak dependence on the companion masses. For low-mass sdBs ($M_\mathrm{sdB}=0.37\,M_\odot$) formed from more massive progenitor stars, the transitional orbital period becomes $\sim \! 0.15\;\mathrm{days}$. These values are very similar to the tidal synchronization boundary ($\sim \! 0.2\;\mathrm{days}$) evident from observations. We discuss the dependence of tidal torques on stellar radii, and we make predictions for the rapidly rotating white dwarfs formed from synchronized sdB binaries.

\end{abstract}


\section{Introduction}

Hot subdwarf B (sdB) stars, first observed by \cite{Humason1947}, are compact and faint stars with surface temperatures between 20,000 and 40,000 K and masses below $0.5\,M_\odot$ \citep{Heber2009,Heber2016,Zhang2009}. These stars have helium-burning cores and thin envelopes \citep{Gotberg2018}, and they are thought to be stripped cores of helium-burning red giants, whose envelopes were previously lost due to some binary interactions \citep{Han2002,Han2003,Pelisoli2020}. About half of the observed sdB systems are found in binaries \citep{Maxted2001,Napiwotzki2004,Geier2011,Copperwheat2011}, with many of them in close ($P_\mathrm{orb}\lesssim10\,\mathrm{d}$) orbits. This suggests that a prior common envelope phase might be responsible for the ejection of their envelopes, as well as the inspirals of their companions to their current orbital configuration \citep{Kruckow2021}.

For binaries with short orbital periods,  tidal interactions can shape both the migration of their orbits and the spin evolution of individual stars. Historically, sdB binaries are sometimes assumed to have reached tidal synchronization, such that their orbital parameters can be derived from measurements of sdB rotation rates (see, e.g., \citealt{Kudritzki1978,Geier2010}), even if the companions (typically white dwarfs or M dwarfs)  are too faint to be seen. However, this assumption has been seriously challenged by observations from the past decade, especially those with high-precision measurements with TESS and {\em Kepler}/K2, where both spin and orbital frequencies are available (see the summary of observation results in Figure \ref{fig:obs_compare}). These studies have found that sdB binaries with orbital periods as short as $\sim \!7$ hours are not always synchronized \citep{Silvotti2022}. Nevertheless, these emerging new data provide an excellent opportunity to test the theoretical modeling of tidal interactions in sdB binaries.

For stars with convective cores and radiative envelopes like sdBs, tidally excited gravity waves in their envelopes are thought to be the most efficient form of tidal interaction \citep{Zahn1977}. These waves are excited by the tidal potential from the companion, and when they propagate through the stellar interior, the fluid damps via radiative diffusion, exerting effective tidal torques that transfer the angular momentum from the orbit to the stellar spin \citep{Zahn1975}. This classical theory of dynamical tides was originally proposed for massive stars, and several works have calculated the tidal evolution of sdB binaries with this model or its adaptations \citep{Geier2010,Pablo2012,Preece2018}.

However, \cite{Ma2023} recently pointed out that an assumption in Zahn's model may not be true for stripped helium-burning stars, like Wolf--Rayet stars in the case of massive stars and sdBs in the case of low-mass stars. While \cite{Zahn1975} assumed that the waves are all efficiently damped when they propagate to the stellar surface, in these stripped stars they may be otherwise reflected and form standing waves. This is particularly true for high-frequency gravity waves excited by short-period orbits, with less efficient radiative damping in stellar envelopes. By direct calculations of tidally excited oscillations with radiative damping, \cite{Ma2023} showed that real tidal torques should have more complicated frequency dependence than the simple power-law relation derived from Zahn's model. This new approach brings concerns to the existing predictions of sdB rotation rates based on Zahn's tidal calculation.

In this paper, we build sdB models and calculate their tidal evolution with the method in \cite{Ma2023}. We carry out direct calculations of stellar oscillations and their tidal response, and we find that standing waves indeed exist in these sdBs. The tidal torques are hence different from Zahn's, and our results for sdB rotation rates are consistent with the observed trends of tidal synchronization for these systems. The manuscript is organized as follows: in Section \ref{sec:physics} we describe the physics of sdB spin-up from tidally excited g-mode oscillations; in Section \ref{sec:model} we describe our model setup; and in  \ref{sec:method_modes} and \ref{sec:method_evolution} we describe how we calculate the oscillation modes and the binary evolution. We show the results for tidal torque calculations in \ref{sec:results_torque} and the spin--orbit evolution of sdB binaries in \ref{sec:results_spin}. We discuss the limitations of our models and the various related physical processes in Section \ref{sec:discussion}. We finally conclude in Section \ref{sec:conclusion}.

\section{Tidal Physics}
\label{sec:physics}

\begin{figure*}
    \centering
    \includegraphics[width=\textwidth]{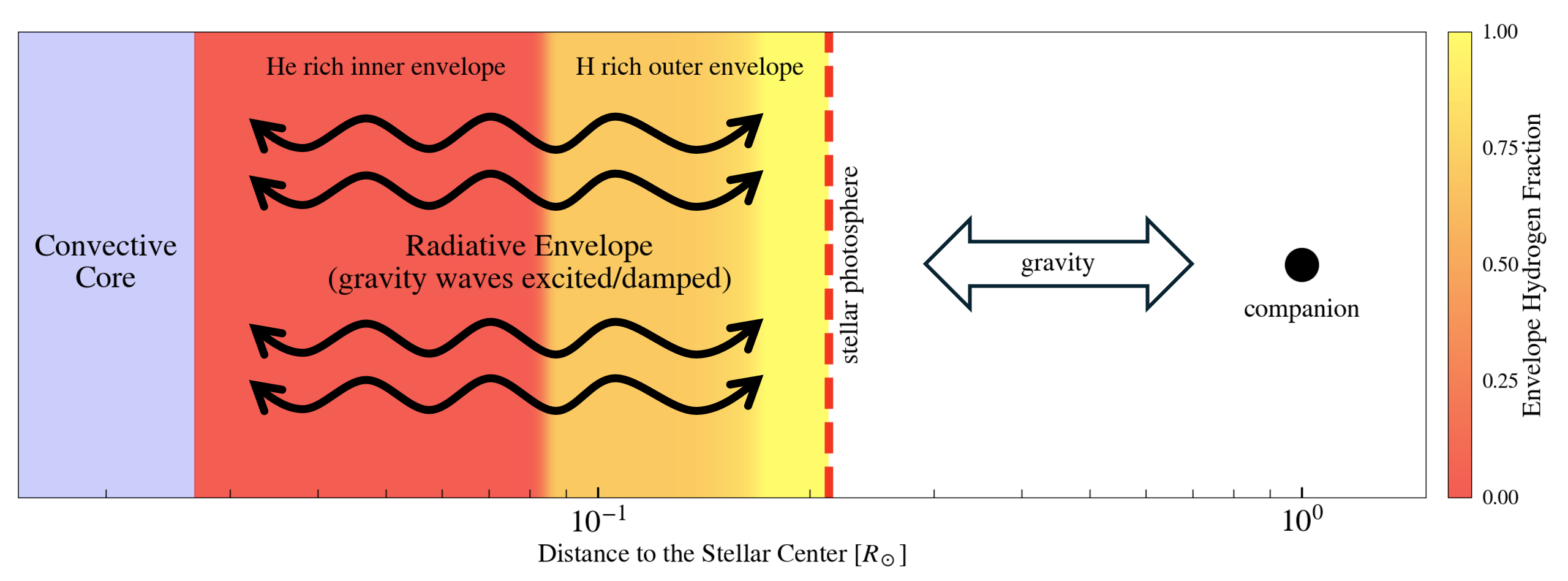}
    \caption{A sketch of the physics of sdB tidal spin-up. Gravity waves, propagating in the radiative envelope, can be tidally excited by the gravity from the orbiting companion. In the hydrogen-rich outer envelope, the waves damp (either partially or fully) and deposit their angular momentum into the star, transferring angular momentum from the orbit to the stellar spin. The color scale shows the hydrogen fraction in the radiative envelope.}
    \label{fig:physics}
\end{figure*}

For subdwarf B binaries, the tidal dissipation inside the sdB star is thought to be responsible for its tidal evolution. In this picture, the tides are excited by the tidal gravity potential from the companion star, which is usually an M--dwarf (dM) or a white dwarf (WD). When the binary orbit is faster than the stellar spin, the tidal dissipation transfers energy and angular momentum from the orbit to the star.

For rotating sdB stars with convective cores and radiative envelopes, three possible tidal dissipation mechanisms could be at work: namely, 1) the turbulent viscous dissipation of equilibrium tidal bulges in the stellar core \citep{Eggleton1998,Eggleton2006}; 2) the turbulent viscous damping of tidally excited inertial waves restored by the Coriolis force \citep{Ogilvie2013,Wu2024}; and, 3) the radiative damping of tidally excited gravity waves in the envelope \citep{Zahn1975,Zahn1977}. Studies have found that the first scenario is usually inefficient for close-in subdwarf binaries, as the orbital periods might be shorter than the convective turnover time in the core, such that convective viscous dissipation is suppressed \citep{Preece2019}. The second scenario is likely unimportant for the same reason, and the small convective core sizes of sdB stars further suppress inertial wave dissipation. We hence focus on the last scenario to be the dominant process for tidal evolution.

We sketch the physics picture of radiative dissipation of tidally excited gravity waves in Figure \ref{fig:physics}. Gravity waves, propagating in the radiative envelope of the star, can be tidally excited by the orbit of the companion. In the outer envelope with large thermal diffusion, these waves damp partially or fully by radiative diffusion, releasing their energy and angular momentum, and hence exert a tidal torque on the star. The orbital angular momentum is thus transferred to the stellar spin. In previous studies, radiative dissipation is often assumed to be efficient so that these gravity waves damp completely in the radiative envelope \citep{Zahn1975}, while in principle they might reflect back and instead form standing waves, i.e. oscillation modes \citep{Ma2023}. Hence, a realistic estimate of tidal torques requires calculation of individual stellar oscillation modes.

For an aligned and circular orbit, the tidal torque for a tidally excited oscillation mode $\alpha$ is given by \citep{Ma2023}:
\beq
\label{eq:tide_torque_mode}
\tau_\alpha = -\frac{m\omega_\alpha \gamma_\alpha q^2M_1R_1^2|W_{lm}Q_\alpha|^2\omega_\mathrm{f}^2}{(\omega_\alpha-\omega_\mathrm{f})^2+\gamma_\alpha^2}\bigg(\frac{R_1}{a}\bigg)^{2(l+1)}\,,
\eeq
where $\omega_\alpha$ and $\gamma_\alpha$ are the mode frequency and growth rate (with $\gamma_\alpha<0$ for damped modes, and the corresponding $\tau_\alpha>0$), and $\omega_\mathrm{f} = m(\Omega_\mathrm{orb}-\Omega_\mathrm{spin})$ is the tidal forcing frequency (measured in the frame co-rotating with the sdB), and $\Omega_\mathrm{spin}$ is the sdB's angular rotation frequency. $M_1$ and $R_1$ are the mass and radius of the sdB, $q=M_2/M_1$ is the mass ratio of the companion to the sdB, $a$ and $\Omega_\mathrm{orb}$ are the semi-major axis and the angular frequency of the orbit. $l$ and $m$ are the mode's angular and azimuthal wave numbers and $W_{lm}$ is an expansion coefficient of the tidal potential. $Q_\alpha\equiv(G/R_1^{l+3})\langle\xi_\alpha|\nabla(r^lY_{lm})\rangle/\omega_\alpha^2$ is the dimensionless overlap integral describing the spatial coupling between the mode and the tidal potential, where $\xi_\alpha$ is the displacement vector of mode $\alpha$, normalized by $\langle\xi_\alpha|\xi_\alpha\rangle\equiv\int_\mathrm{star}\xi_\alpha^*\cdot\xi_\alpha\rho\,dV=M_1R_1^2$. In practice, $Q_\alpha$ is calculated by the relation $Q_\alpha = -(2l+1)\delta\Phi_\alpha/(4\pi\omega_\alpha^2R_1^2)$ \citep{fullerheartbeat:17}, where $\delta\Phi_\alpha$ is the surface gravity potential perturbation. For weak damping ($\gamma_\alpha<\omega_\alpha$), the excitation of individual oscillation modes is independent from each other, such that the total tidal torque can be expressed as
\beq
\label{eq:tide_torque}
\tau_\mathrm{tide}=\sum_\alpha \tau_\alpha\,.
\eeq
Hence, by solving for the internal oscillation modes (with $\omega_\alpha$, $\gamma_\alpha$ and $\mathcal{Q}_\alpha$) inside the sdB, we are able to calculate the torque and the angular momentum transfer rate, given a companion mass and orbit. We stress that there are \textit{no} free parameters in estimating the strength of tidal torques with this method.

\section{Methods}

We calculate the tidal evolution of sdB binaries with the similar method developed for Wolf--Rayet-black hole binaries in \cite{Ma2023}. We first build realistic single evolving sdB models throughout their helium-burning lifetime (Section \ref{sec:model}). We then solve for stellar oscillations based on these models to estimate the tidal torques (Section \ref{sec:method_modes}). Finally, we numerically integrate the coupled spin--orbit evolution of sdB binaries with interpolation between the previously calculated sdB models and tidal torques, with different choices of initial binary parameters, i.e. the initial orbital periods and companion masses (Section \ref{sec:method_evolution}).

\subsection{Stellar Models}
\label{sec:model}

We build single sdB models with the MESA stellar evolution code (r12778; \citealt{Paxton2011,Paxton2013,Paxton2015,Paxton2018,Paxton2019,Jermyn2023}). We build two sdB models to represent the two types of sdBs formed from progenitors of different masses, summarized in Sections \ref{sec:canonical_sdB} and \ref{sec:low-mass_sdB}. We turn on element diffusion for $^1$H, $^4$He, $^{12}$C, $^{14}$N, and $^{16}$O in the MESA models, to account for correct treatments of gravitational settling and radiative acceleration for these atoms. The MESA inlists are available on Zenodo under an open-source 
Creative Commons Attribution license: 
\dataset[doi:10.5281/zenodo.13388526]{https://doi.org/10.5281/zenodo.13388526}.

\subsubsection{$0.47\,M_\odot$ Canonical SdB Model}
\label{sec:canonical_sdB}

This model represents the most abundant sdBs (``canonical'' sdBs) that are formed from low-mass ($M\lesssim2\,M_\odot$) main-sequence progenitor stars. When these stars evolve off main-sequence, they start hydrogen shell burning which deposits helium into their helium core, until they reach the tip of the red giant branch (TRGB) when the helium core exceeds $0.46\,M_\odot$. At this moment, an off-center helium flash is triggered and the helium burning propagates to the center of the core, while the star loses most of its envelope through binary processes (e.g., a common envelope event), leaving a core helium-burning sdB star with a little of its envelope ($\sim \!0.01\,M_\odot$) retained. SdB stars formed this way are insensitive to the masses of their progenitors, and have universal masses of $\approx0.47\,M_\odot$. We establish this sdB model by evolving a $1.2\,M_\odot$ star from zero-age main-sequence (ZAMS) to the TRGB, and then apply an artificial stellar wind to remove its envelope, until the off-center helium burning propagates to the stellar center. This happens at the moment when the envelope mass reaches the desired mass (see details in \ref{sec:env_mass}), due to the specific wind scaling factor we chose. The model then becomes a zero-age canonical sdB model and we evolve it until core helium depletion.

\subsubsection{$0.37\,M_\odot$ Low-mass SdB Model}
\label{sec:low-mass_sdB}

For stars of $\sim \!2-3\,M_\odot$, they can also ignite core helium burning without forming a fully degenerate core, at helium core masses less than $0.46 \, M_\odot$. Hence, they usually form sdBs of lower masses compared to canonical sdBs. To simulate this scenario, we evolve a $2.7\,M_\odot$ star from ZAMS to the TRGB, and then remove its envelope by a similar artificial wind until its envelope mass reaches the desired mass (see details in \ref{sec:env_mass}). The model then triggers central helium burning as a zero-age sdB star. The sdB model we build this way has a mass of $0.37\,M_\odot$.

\subsubsection{Envelope Mass Setup}
\label{sec:env_mass}

SdBs are known to retain a small amount of hydrogen envelope above their helium cores. The amount of hydrogen can be constrained from their spectroscopic properties, and is found to be between $0.001$ -- $0.005\,M_\odot$ (see, e.g., Figure 10 of \citealt{Kupfer2015}). We hence adjust the artificial winds such that the stellar models start core helium burning (zero-age sdB) when they have $10^{-3}\,M_\odot$ hydrogen left. After that moment, we turn off stellar winds as the envelope-stripping phase is considered completed. We note that, real sdB stars can possibly retain more hydrogen than $10^{-3}\,M_\odot$. However, we found many unstable stellar oscillations in sdB models with more massive envelopes, and we are not able to calculate the tidal dissipation of these modes with our current method. We discuss the influence of envelope masses and these unstable modes in more detail in Section \ref{sec:unstable_modes}.

\subsubsection{Convective Core Boundary Setup}

The excitation of gravity waves is sensitive to the size of the convective core, which in turn can be sensitive to how its boundary is treated in stellar evolution models. Unlike the standard convective-overshooting paradigm that has been established for main-sequence stars (see, e.g., the MIST project; \citealt{Choi2016}), overshooting parameters for stars with a helium-burning convective core, like sdBs, are poorly constrained. Nevertheless, asteroseismic measurements of core helium-burning stars suggest the existence of bigger cores compared to theoretical modeling \citep{Constantino2015,Bossini2017,Noll2024}. We hence turn off overshooting for our stellar models in the core helium-burning phase, instead applying the ``predictive mixing'' scheme for convection \citep{Paxton2018}. This allows for a steady growth of the convective core during core helium-burning, more consistent with asteroseismic observations than other choices for convective mixing. Furthermore, the predictive mixing scheme helps prevent ``breathing pulses'' at late stages of the core helium-burning phase, which may split the convective core to create small radiative zones, in which very high-order gravity waves can be trapped. Breathing pulses have been argued to be numerical artifacts \citep{Bauer2021}, and we aim to avoid the associated difficulties in computing gravity modes.

\subsubsection{Rotation Setup}
\label{sec:am_transport}

When stars evolve off main-sequence, their core contracts and spins up, while their envelope expands and spins down. The shear created between the core and the envelope could trigger hydrodynamical and magneto-hydrodynamical instabilities, which transfer some of the core angular momentum to the envelope, forming slowly rotating stellar cores \citep{Fuller2019,Tripathi2024}. Asteroseismic measurements of red clump stars have shown that their core rotation periods are typically $\sim \!100$ days \citep{Mosser2012}. Therefore, if these stellar cores form sdBs, they should also be slowly rotating.

We applied the modified Taylor-Spruit torque as described in \cite{Fuller2019} in our stellar models, and we found that the stellar models at the end of the envelope-stripping phase rotate slowly, with rotation rates insensitive to the initial rotation at ZAMS. The slow rotation rates are consistent with the slow sdB rotation rates measured in wide binaries, where tidal effects are not important (see, e.g., Figure \ref{fig:obs_compare}). We hence set the sdB models to be nonrotating at the start of their helium burning phase. Since we only compute our spin--orbit evolution by post-processing of the stellar models, without actually updating their rotation rates in MESA (see details in \ref{sec:method_evolution}), the single sdB models remain nonrotating throughout their lifetime.

\subsection{Calculation of Oscillation Modes}
\label{sec:method_modes}

We calculate the internal oscillation modes for the individual snapshots of our sdB models with the GYRE stellar oscillation code \citep{townsend:13,townsend2018,goldstein2020}. We solve for non-adiabatic oscillations which account for radiative damping in the oscillation equations. We use the second order Magnus differential scheme, as it proves to be the most reliable when dealing with low-frequency oscillations. We specify our search to $l=m=2$ modes as this is the dominant part of the tidal potential in aligned and circular orbits, with the corresponding $W_{22}=\sqrt{3\pi/10}$. Example GYRE inputs are available on Zenodo under an open-source 
Creative Commons Attribution license: 
\dataset[doi:10.5281/zenodo.13388526]{https://doi.org/10.5281/zenodo.13388526}. When solving for modes, we find that the Brunt-V\"ais\"al\"a frequency \citep{Vaisala1925} profiles solved from MESA are sometimes not consistent with the density and pressure profiles from the same model, which may lead to inaccurate mode solutions. We hence slightly adjust the stellar profiles with the process described in Appendix \ref{app:fix_rho} for our stellar models. We checked that the change of stellar structure due to this process is negligible.

In principle, we need to sum over \textit{all} modes to get the total tidal torque through Equation \ref{eq:tide_torque}. This is not practically possible as there are an infinite number of modes which could be excited at all frequencies. Nevertheless, we note from Equation \ref{eq:tide_torque_mode} that for a given tidal forcing frequency $\omega_\mathrm{f}$, typically only the few nearly resonant modes with $\omega_\alpha$ close to $\omega_\mathrm{f}$ contribute significant torques. Torques from other non-resonant modes are usually negligible due to the $(\omega_\alpha-\omega_\mathrm{f})^2$ term in the denominator of Equation \ref{eq:tide_torque_mode}. We hence restricted our mode solutions to a finite period range, namely from $0.005$ days to $0.5$ days, and we hence found a finite number of modes. We can then calculate the total tidal torques as long as the forcing frequency $\omega_\mathrm{f}$ is between $2\pi/(0.5\,\mathrm{d})=12.57\,\mathrm{d}^{-1}$ and $2\pi/(0.005\,\mathrm{d})=1257\,\mathrm{d}^{-1}$. The $\omega_\mathrm{f}$ calculated from our spin--orbit evolution usually lies well within this range, except for some systems that reach tidal synchronization, whose $\omega_\mathrm{f}$ should approach zero. We hence stop the evolution when $\omega_\mathrm{f}$ reaches the minimum mode frequency $12.57\,\mathrm{d}^{-1}$. We checked that our binary models reaching this condition are at least at 80\% synchronization, so we \textit{define} all systems with $\Omega_\mathrm{spin}\geq0.8\,\Omega_\mathrm{orb}$ as tidally synchronized.

\subsection{Spin--Orbit Evolution}

We integrate the spin--orbit evolution of the sdB binaries from the stellar models and tidal torques we computed. For simplicity, we ignore the tidal dissipation in the companion star (see discussion in Section \ref{sec:diss_caveats}). Assuming circular orbits, the orbital angular momentum of the system is lost due to gravitational wave (GW) radiation and tides, while the sdB receives spin from the tidal torque:
\begin{align}
\label{eq:J_orb_dot}
\dot{J}_\mathrm{orb}&=-\tau_\mathrm{GW}-\tau_\mathrm{tide}\,,\\
\label{eq:J_spin_dot}
\dot{J}_\mathrm{spin}&= \tau_\mathrm{tide}\,,
\end{align}
where $\tau_\mathrm{GW}=(32/5)(G/a)^{7/2} c^{-5} M_1^2M_2^2\sqrt{M_1+M_2}$ is the effective torque by GW radiation \citep{Peters1964}. The GW orbital decay timescale is then given by 
\begin{align}
\label{eq:gw_time}
T_\mathrm{GW}&=5c^5(1+q)^{1/3}P_\mathrm{orb}^{8/3}/(64(4\pi^2)^{4/3}G^{5/3}M_1^{5/3}q)\nonumber \\
&\approx180\,\mathrm{Myr}\,(P_\mathrm{orb}/1\,\mathrm{h})^{8/3}
\end{align}
for an equal mass sdB binary ($q=1$), with a $0.46\,M_\odot$ canonical sdB star, comparable to the sdB lifetime of $\sim \!150\,$Myr for very short-period systems. This means GW orbital decay needs to be included in the spin--orbit evolution.

As we expect efficient angular momentum transport during the core-helium burning phase (\citealt{Fuller2019,Fuller2022}; see discussions in \ref{sec:am_transport}), we assume rigid rotation for the sdB star, with a uniform rotational frequency $\Omega_\mathrm{spin}$. We discuss the case of differential rotation in Section \ref{sec:diff_rot}. The coupled spin--orbit evolution can then be integrated by:
\begin{align}
\label{eq:Omega_spin_dot}
\dot{\Omega}_\mathrm{spin}&=\frac{\dot{J}_\mathrm{spin}}{I_\mathrm{spin}}-{\Omega}_\mathrm{spin}\frac{\dot{I}_\mathrm{spin}}{I_\mathrm{spin}}\,,\\
\label{eq:Omega_orb_dot}
\dot{\Omega}_\mathrm{orb}&=\frac{\dot{J}_\mathrm{orb}}{I_\mathrm{orb}}-{\Omega}_\mathrm{orb}\frac{\dot{I}_\mathrm{orb}}{I_\mathrm{orb}}=-3\frac{\dot{J}_\mathrm{orb}}{I_\mathrm{orb}}\,,
\end{align}
where $I_\mathrm{spin}$ is the moment of inertia of the sdB star, $I_\mathrm{orb}=\mu a^2$ is the moment of inertia of the orbit and we made use of Kepler's Third Law to simplify Equation \ref{eq:Omega_orb_dot}. This means the spin of the sdB star may also change due to the changes of its internal structure and hence moment of inertia.

We make use of the integration machinery constructed in \cite{Ma2023}, with the same interpolation method. As sdB lifetime is typically longer than the Wolf--Rayet stars in \cite{Ma2023}, we choose the integration timestep to be $0.1$ times the values derived from the timestep control method described in \cite{Ma2023}. We integrate the evolution from 1 year after the start of the sdB helium-burning phase, and we stop when the model depletes its core helium (defined by the time when the core helium fraction drops below 1\%) or when the system reaches $\omega_\mathrm{f}\lesssim12.57\,\mathrm{d}^{-1}$ (see Section \ref{sec:method_modes}). The initial rotational period of sdBs is set to be 60 days, to match the observed values from single sdB stars \citep{Silvotti2022}. While single sdBs may not represent a fair sample of binary sdBs at birth, this assumed initial rotational frequency is very low and never important for systems that reach synchronization. As there are only limited theoretical constraints on the initial binary parameters of sdBs, we vary the companion masses uniformly between $0.1\,M_\odot$ and $0.8\,M_\odot$, and choose the initial orbital periods to be between 1 to 18 hours, aiming to cover the observed parameter space of close-in sdB binaries \citep{Schaffenroth2022}. We checked our results are robust against different choices of timestep resolution.

\label{sec:method_evolution}
\section{Results}

In this section, we show the results of our tidal torque calculation and spin--orbit evolution.
\subsection{Tidal Torque}
\label{sec:results_torque}
\begin{figure*}
    \centering
    \includegraphics[width=\textwidth]{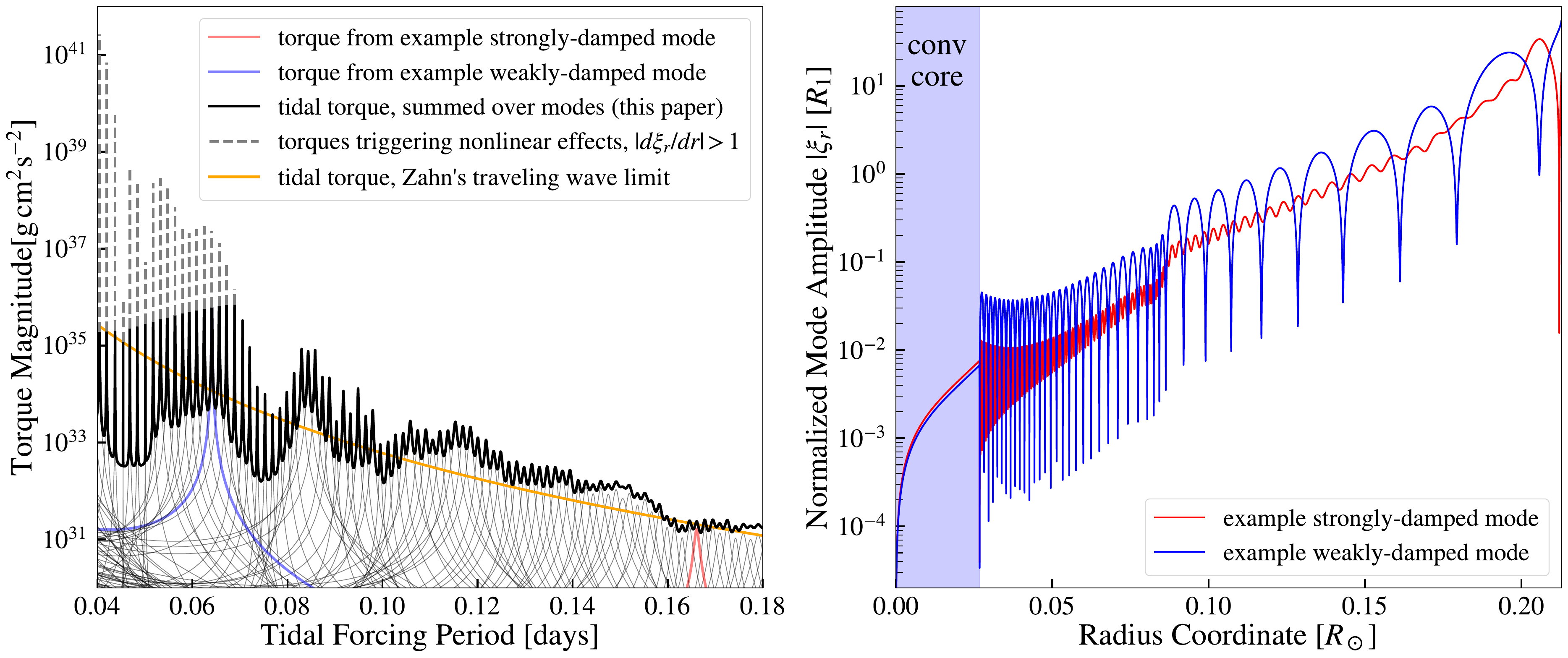}
    \caption{\textbf{Left}: The tidal torque calculated for a $0.47\,M_\odot$ sdB model with a companion of $0.4\,M_\odot$,  when the central helium fraction is 60\%, assuming the sdB is nonrotating. The thick lines show the total torque calculated by summing over contributions from individual tidally excited g modes (thin lines). We also show the torque calculated from Zahn's formalism for comparison. At short periods, the total tidal torque is dominated by resonance peaks from individual weakly-damped g modes, different from the power-law dependence on forcing period of Zahn's formalism. However, on-resonance torques at short periods (thick dashed grey lines) may trigger nonlinear dissipation, and the real torques may not be as large as seen in the plot. At longer periods, g modes are more efficiently damped and the torque agrees better with Zahn's model. Right: the mode eigenfunctions for an example weakly-damped standing mode (blue line) and an example strongly-damped mode (red line) in the left panel. The weakly-damped mode has nodes in its eigenfunction, while the strongly-damped mode efficiently damps near the surface.}
    \label{fig:torque}
\end{figure*}

In Figure \ref{fig:torque}, we show our calculated tidal torque magnitude with its dependence on the period of tidal forcing ($P_\mathrm{f}\equiv 2\pi/\omega_\mathrm{f}$). The calculation is based on the $0.47\,M_\odot$ sdB model with a companion of $0.4\,M_\odot$, when its central helium fraction drops to 60\%, and we assume the sdB is nonrotating such that $\omega_f=m\Omega_\mathrm{orb}$. The torque generally has a complicated dependence on the forcing period. By plotting the torque contributions from each individual oscillation mode $J_\alpha$, we see that this dependence is caused by summing over the resonance peaks of many modes with different frequencies. When the tidal forcing frequency gets close to one of the mode frequencies, the $(\omega_\alpha-\omega_f)^2$ term in Equation \ref{eq:tide_torque_mode} vanishes, and the total torque becomes dominated by the strong resonance peak of that mode. Therefore, the frequencies/periods of these peaks are the frequencies/periods of individual oscillation modes inside the star. These peaks have a nearly uniform period spacing, a feature expected for gravity (g) modes. In the right panel of Figure \ref{fig:torque}, we show some example eigenfunctions of these modes, and we can see that they are indeed g modes that propagate inside the radiative envelope of the star.

Previous studies involving the tidal dissipation of gravity waves in sdBs usually use Zahn's model for dynamical tides \citep{Geier2010,Pablo2012,Preece2018}, which assumes these waves are efficiently damped as they reach the stellar surface (``traveling-wave limit''). We see from the right panel of Figure \ref{fig:torque} that this is clearly not always the case for individual resolved stellar oscillations. The blue line shows the eigenfunction of an example oscillation at short ($P_\mathrm{\alpha}\lesssim 0.14\,\mathrm{days}$) period. We see that instead of efficiently damping near the stellar surface, the wave reflects back at the stellar surface and forms a standing wave with nodes. This means Zahn's picture may overestimate the mode damping rate, hence the tidal dissipation.

For comparison, we show the tidal torques calculated based on Zahn's formalism with a modified formula given by \cite{Kushnir2017}:
\beq
\label{eq:zahn_torque}
\tau_\mathrm{Zahn}=\beta_2\frac{GM_2^2}{r_\mathrm{c}}\bigg(\frac{r_\mathrm{c}}{a}\bigg)^6s_\mathrm{c}^{8/3}\frac{\rho_\mathrm{c}}{\bar{\rho}_\mathrm{c}}\bigg(1-\frac{\rho_\mathrm{c}}{\bar{\rho}_\mathrm{c}}\bigg)^2\,,
\eeq
where $s_\mathrm{c}=\sqrt{3/(\pi G\bar{\rho}_\mathrm{c})}|\Omega_\mathrm{orb}-\Omega_\mathrm{spin}|$, while $r_\mathrm{c},\rho_\mathrm{c}$ and $\bar{\rho}_\mathrm{c}$ are the convective core radius, the density at the core boundary, and the average density of the core, respectively. $\beta_2$ is a dimensionless coefficient solved from stellar structures, and different main-sequence and Wolf--Rayet stellar models have $\beta_2\approx 1$ \citep{Kushnir2017}. Its dependence on the tidal forcing period is mostly the power-law term in $s_c^{8/3}a^{-6}$, as seen in Figure \ref{fig:torque}. This is clearly different from the resonance peak dependence we find from realistic mode calculations. We see that at short periods, if the binary orbit has an off-resonance tidal forcing period (i.e., not close to any stellar oscillation modes), the real tidal torque can be orders of magnitude lower than Zahn's prediction. On the other hand, if the orbit is on resonance, the torque might be significantly larger than Zahn's prediction.

However, when the tidal forcing is on resonance with one of the oscillation modes, the mode amplitude becomes so large that it can trigger nonlinear wave dissipation. In this scenario, the oscillation mode excites a number of nearby daughter and granddaughter modes, and the overall damping rate by this sea of coupled modes could be much larger than the radiative damping of individual modes \citep{Barker2011,Weinberg2012}. The level of nonlinearity can be estimated by the second-order nonlinear term in the momentum equation $\xi\cdot\nabla\xi\sim(d\xi_r/dr)\xi$: when $d\xi_r/dr>1$, nonlinear effects become very strong.

With the above criterion, we showed where tidally excited modes are strongly nonlinear in Figure \ref{fig:torque} with dashed grey lines. We see that nonlinear effects mostly affect on-resonance torques at short ($P_\mathrm{tide}<0.07\,\mathrm{days}$) periods. For these torques, the resonance peaks will be smoothed out when additional nonlinear damping is present, and their actual magnitude may not be as large as seen in Figure \ref{fig:torque}.

As the star and the orbit evolves, both the oscillation mode period (hence the location of the resonance peaks) and the tidal forcing period change over time, so the system can quickly pass through resonances (as long as resonance locking does not happen, see discussions in Section \ref{dis:resonance_lock}). Since the resonances are narrow, the system spends more time out of resonance than in resonance (i.e., with torques much weaker than Zahn's prediction), the accumulated angular momentum received by the sdB star can still be less than the predictions from Zahn's theory.

At longer periods, gravity waves have larger wave numbers, and are expected to damp more efficiently with radiative damping. With larger $\gamma_\alpha$, the $\gamma_\alpha^2$ term becomes more important in the denominator of Equation \ref{eq:tide_torque_mode}, smoothing out the resonance peaks. We see in Figure \ref{fig:torque} that this is exactly the case for the tidal torques at long period ($P_\mathrm{f}>0.14\,\mathrm{d}$), when the individual modes damp so much that the resonance structure gets smoothed out. In addition, the shape of the example eigenfunction (red line) shown in the right panel of Figure \ref{fig:torque} becomes closer to a traveling wave that efficiently damps near the stellar surface, as Zahn's formalism assumes. The tidal torque's dependence on tidal forcing period also gets closer to Zahn's power-law dependence as expected. This further shows that Zahn's traveling wave picture is a limit case of realistic tidal torques at long periods.

\begin{figure*}
    \centering
    \includegraphics[width=0.8\textwidth]{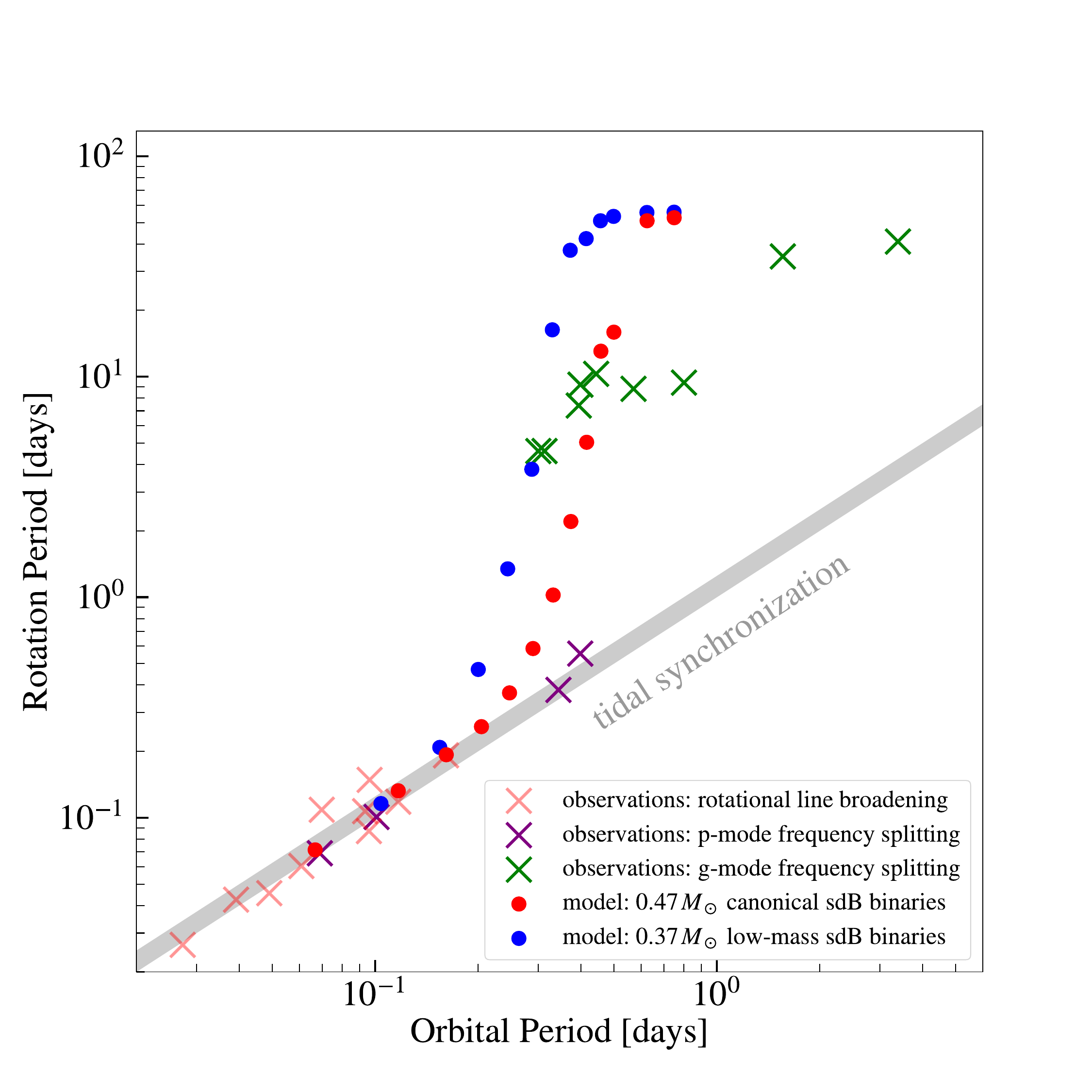}
    \caption{The observed trend of tidal synchronization (defined by $0.8\leq\Omega_\mathrm{spin}/\Omega_\mathrm{orb}\leq1$, shaded grey region) for short-period sdB binaries (crosses) versus the calculation results from binary spin--orbit evolution (dots). The red, purple and green crosses indicate rotational measurements from spectral line broadening, p-mode frequency splitting and g-mode frequency splitting. The red and blue dots indicate the modeled sdB binaries with the $0.47\,M_\odot$ and the $0.37\,M_\odot$ sdB primary, respectively. All modeled binaries have a $0.4\,M_\odot$ companion and initial orbital periods ranging from 1 to 18 hours. For $0.47\,M_\odot$ sdB binaries, we see that all systems with orbital periods less than $\sim \!0.2$ days reach tidal synchronization, while for $0.37\,M_\odot$ sdB binaries, this synchronization period becomes $\sim \!0.15$ days due to weaker torques on these smaller sdBs. These results match the observed trends of sdB tidal synchronization. In addition, most systems with $0.3\,\mathrm{d}\lesssim P_\mathrm{orb}\lesssim 0.6\,\mathrm{d}$ are spun-up to rotational periods of a few days, which also agree with the observed period range of partially synchronized binaries.}
    \label{fig:obs_compare}
\end{figure*}

\subsection{Tidal Synchronization}
\label{sec:results_spin}

With the tidal torques calculated, we are able to integrate the coupled spin--orbit evolution of our models. In Figure \ref{fig:obs_compare}, we show the calculation for our $0.47\,M_\odot$ canonical sdB model and $0.37\,M_\odot$ low-mass sdB model with a fixed companion mass of $0.4\,M_\odot$, with different initial orbital periods. The rotational and orbital periods are evaluated at the end of the spin--orbit evolution (defined in Section) \ref{sec:method_evolution}, and some ultra-short-orbit synchronized systems that reaches $P_\mathrm{orb}<0.02\,\mathrm{d}$ due to gravitational wave orbital decay are not shown. We see that for the $0.47\,M_\odot$ sdB model, all systems with orbital periods less than $\sim \! 0.2$ days reach tidal synchronization, while for the $0.37\,M_\odot$ sdB model, the synchronization period becomes $\sim \! 0.15$ days.

To compare with observations, we also plot the measured rotational and orbital periods for short-period sdB binaries in Figure \ref{fig:obs_compare}. The different colors show the sdB rotation rates derived from spectral line measurements (HS 0705+6700, \citealt{Drechsel2001}; CD-30 11223, \citealt{Vennes2012}; SDSS J162256.66+473051.1, \citealt{Schaffenroth2014}; PTF1 J0823+0819, \citealt{Kupfer2017}; PTF1 J011339.09+225739.1, \citealt{Wolz2018}; ZTF J2130+4420, \citealt{Kupfer2020}; ZTF J2055+4651, \citealt{Kupfer2020b}; SDSS J082053.53+000843.4, \citealt{Schaffenroth2021}; HW Vir, \citealt{Esmer2021}; and EPIC 216747137, \citealt{Silvotti2021}), asteroseismic p-mode frequency splitting (NY Vir, \citealt{Charpinet2008}; Feige 48, \citealt{VanGrootel2008}; V1405 Ori, \citealt{Reed2020}; and HD 265435, \citealt{Pelisoli2021}), or g-mode frequency splitting (KIC 11179657 and KIC 2991403, \citealt{Pablo2012}; FBS 1903+432, \citealt{Telting2014}; KIC 7664467, \citealt{Baran2016}; EQ Psc and PHL 457, \citealt{Baran2019}; KIC 2438324, \citealt{Sanjayan2022}; TYC1 4544-2658-1, \citealt{Silvotti2022}; and PG 0101+039, \citealt{MaXY2023}), respectively.

We see that all the observed systems with $P_\mathrm{orb}\lesssim 0.2\,\mathrm{d}$ are close to tidal synchronization, while all but two systems\footnote{
The two exceptional systems are Feige 48 and V1405 Ori. For Feige 48, there are some discrepancies on its orbital and rotational periods measured (see, e.g., \citealt{Fontaine2014,Preece2018,Baran2024}). For V1405 Ori, there are some evidences that it might a differentially rotating sdB \citep{Reed2020}, such that its envelope can be synchronized at longer periods while its interior is not (see discussions in Section \ref{sec:diff_rot}).} above this period are not synchronized. This matches strikingly well with the theoretical prediction from our sdB models. In addition, the models with $0.3\,\mathrm{d}\lesssim P_\mathrm{orb}\lesssim 0.6\,\mathrm{d}$ are tidally spun-up to a rotational period of a few days, also consistent with the observed partially-synchronized systems in that period range. Hence, our theoretical calculations agree with the observation data.

\begin{figure*}
    \centering
    \includegraphics[width=\textwidth]{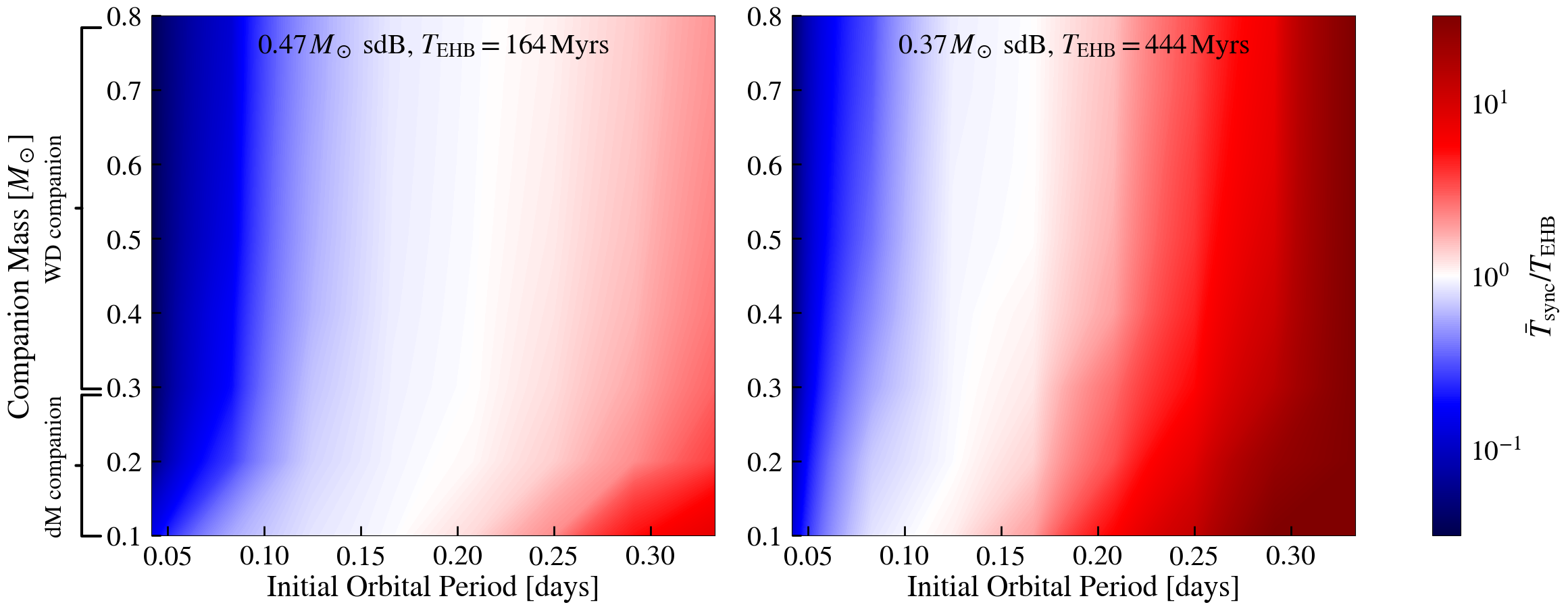}
    \caption{The ratio between the tidal synchronization timescale $\bar{T}_\mathrm{sync}$ and the sdB lifetime $T_\mathrm{EHB}$ interpolated between different choices of companion masses and initial orbital periods. The blue regions show the parameter space where $\bar{T}_\mathrm{sync}<T_\mathrm{EHB}$. The brackets on the $y$ axis indicate the typical companion masses for sdB+WD or sdB+dM systems \protect\citep{Schaffenroth2022}. \textbf{Left}: The results for the $0.47\,M_\odot$ sdB model, with $T_\mathrm{EHB}=164\,\mathrm{Myrs}$. We see that for systems in orbits less than $\sim \! 0.15 - 0.22\,\mathrm{d}$, the synchronization timescale is less than the stellar lifetime, meaning these systems are likely to be observed as tidally synchronized. The results have weak dependence on companion masses. \textbf{Right}: The results for the $0.37\,M_\odot$ sdB model. The critical orbital period below which systems become synchronized now becomes $\sim \! 0.10 - 0.17\,\mathrm{d}$.}
    \label{fig:sync_timescale}
\end{figure*}

Note that in Figure \ref{fig:obs_compare} the spin and orbital frequencies are shown at the end of the spin--orbit evolution, while measurements for realistic systems usually occur when the sdB still undergoes core-helium burning. Therefore, it is illustrative to show the tidal synchronization timescales calculated for our models, and to compare them with the sdB lifetime $T_\mathrm{EHB}$. If the synchronization time is shorter, then we expect those systems are likely to reach tidal synchronization to be observed. As we consider systems with 80\% synchronization as synchronized (see discussions in Section \ref{sec:method_modes}), we define the tidal synchronization timescale throughout the whole evolution as:
\beq
\label{eq:sync_time}
\bar{T}_\mathrm{sync}\equiv \left\{
\begin{array}{ll}
      t_{\Omega_\mathrm{spin}/\Omega_\mathrm{orb}=0.8}\, & ,\;\mathrm{if\;synchronized}\, \vspace{7pt}\\
      0.8\,T_\mathrm{EHB}\, \bigg(\frac{\Omega_\mathrm{spin}}{\Omega_\mathrm{orb}}\bigg)^{-1}_\mathrm{final}\, &,\; \mathrm{if\;not}\, \\
\end{array} 
\right. 
\eeq
where $t$ is the stellar age since the start of sdB core helium-burning. We then run a grid of spin--orbit evolution for both of our $0.47\,M_\odot$ and $0.37\,M_\odot$ sdB models, with initial orbital periods of $(1,2,3,4,5,6,7,8)$ hours and companion masses of $(0.1,0.2,0.3,0.4,0.5,0.6,0.7,0.8)$ solar-masses, and calculated their tidal synchronization timescale defined by Equation \ref{eq:sync_time}.

We show the interpolated results for the ratio $\bar{T}_\mathrm{sync}/T_\mathrm{EHB}$ based on this grid in the whole parameter space in Figure \ref{fig:sync_timescale}. On the $y$-axis we also label the typical companion masses for sdB+dM and sdB+WD binaries \citep{Schaffenroth2022}. We see that this ratio ranges from $10^{-1.5}$ to $10^{1.5}$, with weak dependence on the masses of the companion star. This is an expected result from our tidal torque formula (Equation \ref{eq:tide_torque_mode}): the torque depends on the mass-ratio (hence the secondary mass) as $\tau_\alpha\propto q^2a^{-6}$, where $a=(G(M_1+M_2)/\Omega_\mathrm{orb}^2)^{1/3}\propto(1+q)^{1/3}$, hence $\tau_\alpha\propto q^2/(1+q)^2$. For typical sdB binaries, $q$ ranges from 0.3 to 1.5, and the corresponding torque scaling is maximally different by only a factor of $\sim \! 7$.
In contrast, $\bar{T}_\mathrm{sync}/T_\mathrm{EHB}$ varies by a factor of $\sim \! 10^3$ over the period range shown in Figure \ref{fig:sync_timescale}, due to its strong dependence on semi-major axis.
 
The blue colored regions in Figure \ref{fig:sync_timescale} show the parameter space where $\bar{T}_\mathrm{sync}/T_\mathrm{EHB}<1$, or where the binaries are expected to be synchronized. We see that for $0.47\,M_\odot$ sdB binaries, systems with initial orbital periods less than $\sim \! 0.15 - 0.22$ days have synchronization timescales shorter than the 164 Myrs sdB lifetime, while for $0.37\,M_\odot$ sdB this period becomes $\sim \!0.10 - 0.17$ days for its 444 Myrs lifetime, depending on the companion masses. As binaries at these orbital periods produce weak GW emission, their orbital periods are nearly constant, and these results confirm the critical orbital periods for synchronization shown in Figure \ref{fig:obs_compare}.

We note from Figure \ref{fig:sync_timescale} that, even though systems below the synchronization periods can reach tidal synchronization in the sdB lifetime, in most of the parameter space their synchronization timescales are not less than the corresponding sdB lifetime by one order of magnitude. This is especially true for sdB+dM binaries with $M_\mathrm{companion}\lesssim0.3\,M_\odot$, and we can see that $\bar{T}_\mathrm{sync}<0.1\,T_\mathrm{EHB}$ is only achieved for those binaries in $P_\mathrm{orb}\lesssim0.05\,\mathrm{d}\approx1\,\mathrm{hour}$ orbits. This is consistent with the findings that sdB binaries with small companions (dMs or brown dwarfs) can be slightly sub-synchronized even at orbital periods less than $\sim \!2.5$ hours (e.g., SDSS J162256.66+473051.1, a 64\% synchronized system with $P_\mathrm{orb}=1.67\,\mathrm{h}$, \citealt{Schaffenroth2014}; and SDSS J082053.53+000843.4, a 65\% synchronized system with $P_\mathrm{orb}=2.3\,\mathrm{h}$, \citealt{Schaffenroth2021}). 

\cite{Schaffenroth2021} further point out that synchronized binaries locate further away from the zero-age extreme horizontal branch (ZAEHB) on the $\log g-T_\mathrm{eff}$ diagram compared to these sub-synchronized systems, suggesting that those synchronized binaries might be older. Our findings that the tidal synchronization timescales at small orbital periods are shorter than the sdB lifetime (but not by orders of magnitude), agrees with this explanation.

Historically, the orbital inclinations and companion masses are hard to acquire for non-eclipsing sdB binaries. Some works hence assume tidal synchronization for short-period binaries, and derive the orbital parameters from the orbital periods by setting $P_\mathrm{orb}=P_\mathrm{rot}$ (e.g., \citealt{Geier2010}). However, if we can assume tidal synchronization for systems with $\bar{T}_\mathrm{sync}<0.1\,T_\mathrm{EHB}$, we see that this method should only apply to binaries with $P_\mathrm{orb}\lesssim1\,\mathrm{h}$. This is much shorter than the synchronization period ($P_\mathrm{orb,\,sync}=1.2\,\mathrm{d}$) assumed in \cite{Geier2010}, meaning that in their work  the companion masses/inclinations might be over/underestimated.

Additionally, binaries with $P_\mathrm{orb}\lesssim1\,\mathrm{h}$ can undergo significant orbital decay due to gravitational-wave radiation, and it is questionable whether these systems can \textit{ever} reach 100\% tidal synchronization, as tides at sub-synchronization may not be strong enough for $\Omega_\mathrm{spin}$ to fully catch up with $\Omega_\mathrm{orb}$ (as in the case of WD binaries, see, e.g., \citealt{Scherbak2024}). Mass-transfer may also happen for these binaries, making their evolution more complicated \citep{Bauer2021}.

\section{Discussion}
\label{sec:discussion}

\subsection{Tidal Torque Scaling with Stellar Radii}
\label{sec:unstable_modes}

\begin{figure*}
    \centering
    \includegraphics[width=\textwidth]{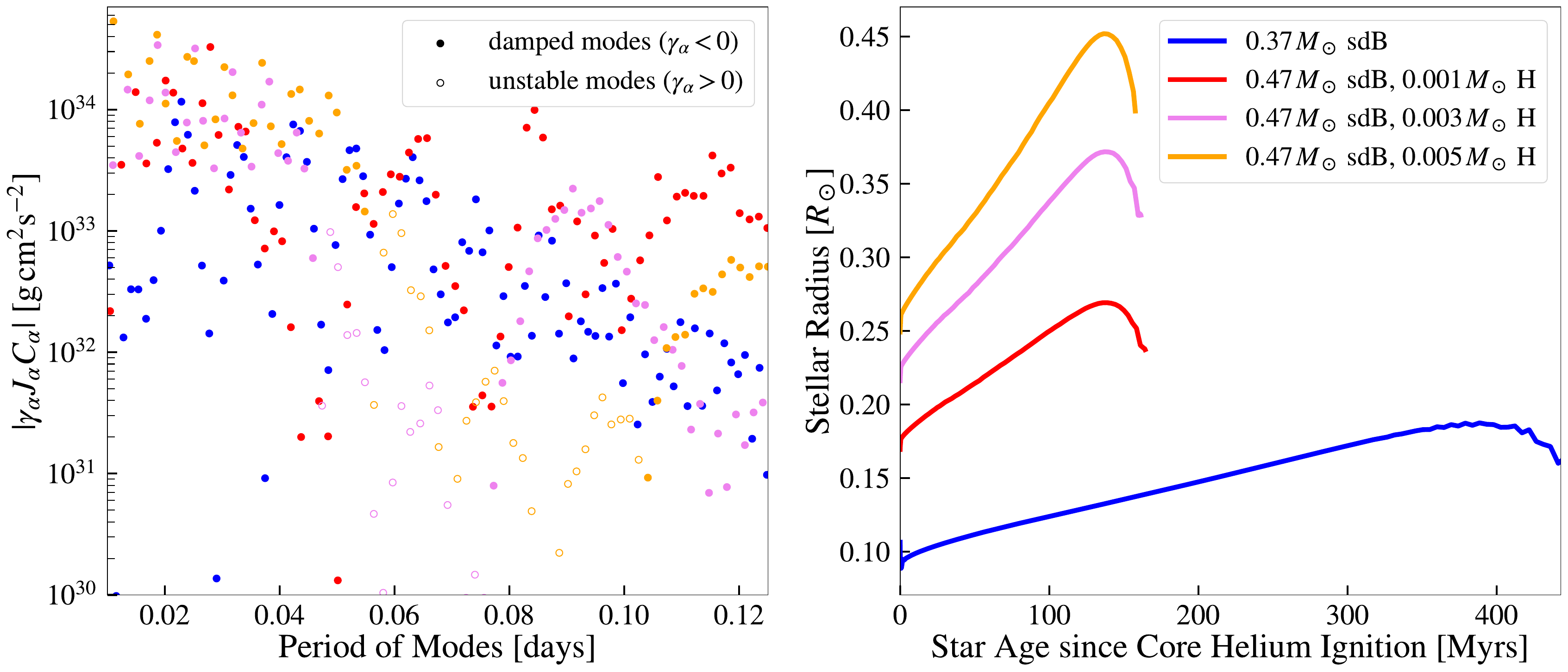}
    \caption{\textbf{Left}: The magnitude of $\gamma_\alpha J_\alpha C_\alpha$ (defined in the main text) for different sdB models. This quantity determines the tidal torque of each mode per unit tidal forcing strength, and we see that they are of similar orders of magnitude without clear dependence on the different stellar models. Hence, the physical torque should be roughly proportional to the tidal forcing strength which scales as $R^6$. Note that sdB models with hydrogen masses greater than $10^{-3}\,M_\odot$ have some unstable modes, and their tidal excitation can not be treated with our current method. \textbf{Right}: Stellar radii as a function of age for different sdB models. More massive systems with more hydrogen left in the envelope have larger radii, and the tidal torques are expected to be stronger based on the $R^6$ scaling.}
    \label{fig:damping_compare}
\end{figure*}

We saw in Section \ref{sec:results_spin} that the binary orbital period required to reach tidal synchronization is shorter for the $0.37\,M_\odot$ low-mass sdB, compared to the $0.47\,M_\odot$ canonical sdB. This means the tidal torque must be weaker for low-mass sdBs. To explain the reason, we consider an equal-mass binary ($q=1$), and rewrite the mode torque of equation \ref{eq:tide_torque_mode} as
\beq
\tau_\alpha = -f(\omega_\mathrm{f})\gamma_\alpha J_\alpha C_\alpha S(R_1)\,,
\eeq
where
\beq
f(\omega_\mathrm{f})\equiv\frac{\omega_\mathrm{f}^2}{(\omega_\alpha-\omega_\mathrm{f})^2+\gamma_\alpha^2}
\eeq
is a dimensionless function describing the resonance dependence of the torque on $\omega_\mathrm{f}$, whose scaling should be of similar order for different stellar models. The quantity
\beq
J_\alpha\equiv m\omega_\alpha  \langle\xi_\alpha|\xi_\alpha\rangle 
\eeq
is the angular momentum of the oscillation mode $\alpha$ with azimuthal wave number $m$, which roughly scales as $J_\alpha\propto M_1 R_1^2$ for stars of similar structure at the same mode frequency. For dissipating modes, the rate for mode $\alpha$ to dissipate its angular momentum (i.e., exerting a torque) is $\gamma_\alpha J_\alpha$. The quantity
\beq
C_\alpha\equiv\frac{|W_{lm}Q_\alpha|^2}{\langle\xi_\alpha|\xi_\alpha\rangle/M_1R_1^2}
\eeq
describes the dimensionless coupling of the tidal potential and the oscillation mode, and $S(R_1) \equiv (R_1/a)^{2(l+1)}$ is the scaling of the tidal forcing strength.

With this notation, we can see that the quantity $\gamma_\alpha J_\alpha C_\alpha$ represents the rate at which mode $\alpha$ deposits its angular momentum into the star (i.e., the tidal torque), per tidal forcing strength. We plot this quantity for the $0.47\,M_\odot$ and $0.37\,M_\odot$ sdB models in Figure \ref{fig:damping_compare}, and we see that they have similar orders of magnitude without clear dependence on the different stellar models. This is expected as these sdB stars have very similar internal structures.

Therefore, the main scaling of the physical tidal torque comes from the forcing strength $S(R_1)$. As shown in right panel of Figure \ref{fig:damping_compare}, the $0.37\,M_\odot$ sdB is more compact than the $0.47\,M_\odot$ sdB due to its lower mass, and its radius $R_1$ is smaller by a factor of $\sim \! 2$. For $l=2$ modes, this produces a difference in the tidal torque by $ (R_{0.37}/R_{0.47})^6 \sim  1/64$. Since the $0.37\,M_\odot$ sdB has a smaller moment of inertia by a factor of a few, its synchronization time scale is about ten times longer. We can confirm this result from Figure \ref{fig:sync_timescale}: the tidal synchronization timescale at $P\sim \!0.2\,\mathrm{d}$ for $0.47\,M_\odot$ sdB binaries is $T_\mathrm{EHB,0.47}=164$ Myrs, roughly 10 times shorter than the timescale of $0.37\,M_\odot$ sdB binaries at the same period (a few times $T_\mathrm{EHB,0.37}=444$ Myrs).

The above analysis can also provide insight into the tidal torques for sdBs with different hydrogen envelope masses. Detailed sdB modeling has shown that hydrogen envelope masses range from $0.001$ to $0.005\,M_\odot$ \citep{Kupfer2015}. This small amount of hydrogen never affects the core structure of the helium-burning sdBs, but it can greatly change the stellar radius.

In the right panel of Figure \ref{fig:damping_compare} we show the stellar radii for some $0.47\,M_\odot$ sdB models that retained more hydrogen than $10^{-3}\,M_\odot$. Compared to the original $10^{-3}\,M_\odot$ hydrogen model, we can see that even a slight increase of hydrogen could increase the sdB radius by a factor of $\sim \! 1.5-2$. We further plot the $\gamma_\alpha J_\alpha C_\alpha$ calculated for these models in the left panel of Figure \ref{fig:damping_compare}, and we see that despite some scatter, they are similar to the $10^{-3}\,M_\odot$ hydrogen model. We hence expect the $\tau\propto R^6$ scaling roughly holds for these models, and the tidal torque for these larger sdBs could be larger by a factor of $\sim \! 10-50$, which will increase the synchronization transitional period.

We note, however, there is a reason that we did not actually compute the tidal torques for these more extensive sdB models. We see in Figure \ref{fig:damping_compare} that for sdB models with $M_\mathrm{H}>10^{-3}\,M_\odot$, there exists a period range where the mode growth rate $\gamma_\alpha$ is positive, i.e. where the modes are {\it unstable}. This is caused by the so-called $\kappa$-mechanism in these stars, where the partial ionization of iron creates an opacity bump, generating self-excited oscillations \citep{Charpinet1996,Charpinet1997}. Our torque in equation \ref{eq:tide_torque_mode} only holds for damped oscillation modes, and it is unclear how these self-excited oscillations would interact with tidal forcing (see, e.g., \citealt{Fuller2021}). These unstable modes have periods of $\sim \! 0.05-0.1$ days, so they could be very important for the tidal evolution of sdBs in $\sim \! 0.1-0.2$ day orbits. Future works should investigate how these modes will behave under tidal excitation.

The above analysis also explains the discrepancies between the tidal synchronization period we calculated and those estimated by \cite{Preece2018}, who applied Zahn's traveling wave limit. The synchronization periods we found for the $0.47\,M_\odot$ canonical sdB model are longer than the periods from their work, which means the tidal torque in our cases is stronger. This might be because \cite{Preece2018} used an sdB model with only $10^{-4}\,M_\odot$ hydrogen left, whose radius is smaller than our model. Hence, even though the gravity waves are more damped with Zahn's traveling wave limit in their models, the tidal torque can still be weaker due to its strong dependence on the stellar radius.

\subsection{Resonance Locking}
\label{dis:resonance_lock}

In binary systems, if the tidal torque consists of many resonance peaks from individual modes, a process called resonance locking may occur \citep{Witte1999,Witte2001}. In this scenario, the forcing frequency of the binary enters a resonance with one of the oscillation modes $\alpha$, and stays as
\beq
\omega_\mathrm{f} \equiv m(\Omega_\mathrm{orb}-\Omega_\mathrm{spin}) \simeq \omega_\alpha
\eeq
throughout the binary lifetime. As $\omega_\alpha$ evolves on its own timescale which is independent of the binary separation, this scenario may result in very different binary evolution history compared to other tidal theories.

To see whether resonance locking can happen for sdB binaries, we write the evolution of forcing frequency as
\beq
\dot{\omega}_\mathrm{f}=m\bigg(\frac{3(\tau_\mathrm{tide}+\tau_\mathrm{GW})}{I_\mathrm{orb}}-\frac{\tau_\mathrm{tide}}{I_\mathrm{spin}}\bigg)\,,
\eeq
where we substitute Equations \ref{eq:J_orb_dot}, \ref{eq:J_spin_dot}, \ref{eq:Omega_spin_dot} and \ref{eq:Omega_orb_dot}, and neglect the $\dot{I}_\mathrm{spin}$ term as the stellar structure and moment of inertia only changes slowly during the evolution. To maintain a resonance lock, we must have $\dot{\omega}_\mathrm{f}=\dot{\omega}_\alpha$. For helium-burning subdwarfs, their g-mode frequency increases over time, hence the necessary (but not sufficient) condition for resonance locking to occur is $\dot{\omega}_\mathrm{f}>0$, or
\beq
1+\frac{\tau_\mathrm{GW}}{\tau_\mathrm{tide}}>\frac{ I_\mathrm{orb}}{3I_\mathrm{spin}}\,.
\eeq

For binaries of order-of-unity mass ratios, $I_\mathrm{orb}=\mu a^2\sim M_1a^2\gg M_1R_1^2>I_\mathrm{spin}$. The above relation hence never holds for realistic sdB binaries, unless $\tau_\mathrm{GW}\gg \tau_\mathrm{tide}$. This can happen either for already close-to-synchronization binaries or for wide binaries, where $\tau_\mathrm{tide}$ becomes very small in both cases. In the former case, the low-frequency oscillating g-modes that contribute most to the tidal torque should be very efficiently damped (see Section \ref{sec:results_torque}), which prevents resonance peaks from forming. In the latter case, tidal evolution is not important, because it would occur on a gravitational wave inspiral time which is very long for wide binaries. We hence do not expect resonance locking to happen for sdB binaries, which is confirmed with our numerical spin--orbit evolution calculations.

\subsection{Differential Rotation}
\label{sec:diff_rot}

As we expect very efficient angular momentum transport inside the sdBs, we assume they are rigidly rotating in our spin--orbit evolution calculations. Observationally, asteroseismology can measure the internal rotation of stars via frequency splittings of g-mode and p-mode oscillations \citep{Aerts2010book}. Since g-modes mainly probe the deeper region of the star, while p-modes probe the outer layers, a difference between the rotational rates derived from their frequency splittings may suggest the level of differential rotation between the stellar core and the outer layers.

With {\it Kepler}/K2, there have now been a handful of pulsating sdBs with both p-mode and g-mode frequency splitting measured. \cite{Kern2018} reports that for the sdB+WD system KIC 11558725, the rotational rate derived from p-mode splitting is $P_\mathrm{p}=40.2\pm 0.3\,\mathrm{days}$, while the rate from g-mode splitting is found to be $P_\mathrm{g}=45.1\pm 7.8\,\mathrm{days}$, showing that KIC 11559725 is roughly rigidly-rotating. Similar results are found for sdBs EPIC 220422705 (\citealt{MaXY2022}; where $P_\mathrm{p}\sim 29\,\mathrm{days}$ and $P_\mathrm{g}\sim 32\,\mathrm{days}$), and PG 0101+039 (\citealt{MaXY2023}; where $P_\mathrm{p}=8.60\pm 0.16\,\mathrm{days}$ and $P_\mathrm{g}=8.81\pm 0.06\,\mathrm{days}$). We note that the slightly faster rotation rates measured from p-mode splitting for these systems are consistent with our tidal spin-up picture with dissipating gravity waves: as gravity waves mostly dissipate in the outer envelopes of the star (Figure \ref{fig:physics}), they exert local tidal force mostly in these regions, and angular momentum is subsequently transported inward.

However, the above systems all have slow rotation rates compared to the typical rotational periods ($P_\mathrm{rot}\lesssim 0.3\,\mathrm{d}$) of tidal synchronization calculated from our models. This means the tidal torque are weak for these systems. We pay particular attention to two systems: 1) V1405 Ori (also EPIC 246683636 or KUV 04421+1416), an sdB+dM binary with an orbital period of 0.498 days \citep{Reed2010}. \cite{Reed2020} obtained the p-mode and g-mode splitting from K2 observations, and determined a p-mode derived rotation rate of $0.555\pm 0.029\,\mathrm{days}$ and  (marginally) a g-mode derived rotation rate of $4.2\pm 0.4\,\mathrm{days}$. If the $P_\mathrm{g}$ measurements are reliable, this appears to be a differentially-rotating sdB whose envelope is almost synchronized while the interior is not; 2) TIC 441725813 (also TYC 4427-1021-1, 2MASS J17045838+7304433, or Gaia DR3 1655107708129775744), a system with recent asteroseismic measurements from TESS. \cite{Su2024} derived a $P_\mathrm{p}=17.9\pm 0.7\,\mathrm{days}$ and $P_\mathrm{g}=85.3\pm 3.6\,\mathrm{days}$ from the data, and their initial analysis showed that it is possibly in a binary orbit of $0.28$ days, slightly longer than the synchronization orbital period we found in theories. If the orbit is confirmed, this will be a sub-synchronized sdB star with substantial differential rotation.

The above systems suggest that for strong tidal torques acting in the sdB outer envelope, the assumption of rigid rotation might break down. As the stellar core, which carries most of the stellar mass and moment of inertia, is weakly coupled to the envelope in this case, our calculated tidal synchronization orbital period might be shorter than that needed to synchronize the envelope.

While the current sample is limited, we expect further observations with TESS can provide us more sdB pulsators with differential rotation measured \citep{Baran2023,Baran2024,Uzundag2024}. It is beyond the scope of this work to develop methods for the spin--orbit evolution of differentially rotating sdB models, but we comment that it might be a crucial factor to understand sdB tidal spin-up.

\subsection{Implications for Rotation Periods of CO WDs}

Rotating sdBs can potentially form rapidly rotating carbon--oxygen (CO) white dwarfs after their nuclear burning stops. If the spin angular momentum is conserved after the core-helium-burning phase, the rotation periods of CO WDs are then given by:
\beq
P_\mathrm{rot,\,WD}=\frac{I_\mathrm{WD}}{I_\mathrm{sdB,\,end}}P_\mathrm{rot,\,end}\,,
\eeq
where $I_\mathrm{sdB,\,end}$ and $P_\mathrm{rot,\,end}$ are the moment of inertia and the rotation period of the sdB at the end of its helium-burning phase (defined by the time when the central helium fraction drops below 1\%), and $I_\mathrm{WD}$ is the moment of inertia of the CO WD. Evolving our $0.47\,M_\odot$ sdB model until it forms a CO WD, we find that the stellar moment of inertia decreases from $I_\mathrm{sdB,\,end}=3\times10^{51}\,\mathrm{g\,cm}^{2}$ to $I_\mathrm{WD}=1.8\times10^{50}\,\mathrm{g\,cm}^{2}$ in 50 Myrs, due to the shrinking of the star after nuclear burning stops.


Since tidal synchronization occurs at orbital periods less than $\approx \! 0.2\, \mathrm{d}$, rapidly rotating CO WDs formed from tidally synchronized sdBs should have $P_\mathrm{rot,\,WD}\lesssim(1.8\times10^{50}/3\times10^{51})\times 0.2\,\mathrm{d}\approx17\,\mathrm{min}$. At orbital periods $P_\mathrm{orb}\lesssim1\,\mathrm{h}$, the gravitational-wave decay timescale becomes less than the sdB lifetime, so the sdB does not form a WD before mass transfer with its companion. This gives a lower-limit of $P_\mathrm{rot,\,WD}\gtrsim(1.8\times10^{50}/3\times10^{51})\times 1\,\mathrm{h}\approx4\,\mathrm{min}$ if mass transfer has not occurred. We hence crudely estimate that rapidly rotating CO WDs formed from synchronized sdB stars can have rotation periods between 4 to 17 minutes. This corresponds to rotation rates roughly a hundred times larger than ordinary WDs.

In the above analysis, we ignored any tidal torques after the core-helium-burning phase. Since the star can rotate much faster than the orbit as it contracts, tidal dissipation may spin it back down, producing longer rotation periods than those listed above. Future works should investigate this scenario to have more realistic estimates of the rotation rates of CO WDs originating from sdB binaries. Nonetheless, future observations of rapidly rotating CO WDs in close binaries may indicate that they were tidally spun up during an sdB phase of evolution.

\subsection{Limitations with the Mode Decomposition Method}
\label{sec:diss_mode_method}

Throughout this work, we calculate the total tidal torque by expanding it into a summation of tidal torques from individual oscillation modes (Equation \ref{eq:tide_torque}). \cite{Townsend2023} recently pointed out a potential issue with this ``mode decomposition'' method. They found that the magnitude of off-resonance torques could be different from the magnitude calculated from direct solving of fluid equations under the same tidal potential. \cite{Dewberry2024} introduced a correction to the method for viscous damping addressing this issue, yet it cannot directly apply to the radiative dissipation of tidally excited g-modes.

Nevertheless, when tidal torques are dominated by resonant modes, the correction to the tidal torque magnitude is likely only significant when the tidal forcing frequency is off-resonance \citep{Townsend2023}. Therefore, our results should still be valid as long as most tidal-spin up is caused by on-resonance torques. Future works should improve our method to investigate whether off-resonance corrections change our results.

We did not include nonlinear dissipation of modes in our calculations (see discussions in Section \ref{sec:results_torque}). There are no theoretical works on estimating this dissipation on core helium burning stars (except for some toy models, see, e.g., \citealt{Ma2023}). Nevertheless, as shown in Figure \ref{fig:torque}, strong nonlinear effects should only affect sdBs in the shortest-period orbits which would be tidally synchronized quickly. We therefore suspect that nonlinear effects do not greatly change the synchronization periods we calculated, which matches the observed trends fairly well. Future works should look into the nonlinear effects and their potential influence on sdB tidal spin-up.

\subsection{Tidal Heating}

Tides not only transfer angular momentum between the star and the orbit, but can also dissipate energy through tidal heating, an effect we ignored in our modeling. The rate of tidal heating can be estimated as:
\beq
\dot{E}\sim\dot{J}(\Omega_\mathrm{orb}-\Omega_\mathrm{spin})=\tau_\mathrm{tide}(\Omega_\mathrm{orb}-\Omega_\mathrm{spin}).
\eeq

For close sdB binaries, $\Omega_\mathrm{orb}\sim2\pi/0.2\,\mathrm{days}\approx0.4\,\mathrm{mHz}$, and the strongest tidal heating occurs when $\Omega_\mathrm{spin}=0$. Therefore, off-resonance tidal torques between $10^{32}-10^{35}\,\mathrm{g\,cm}^2\mathrm{s}^{-2}$ (see Figure \ref{fig:torque}) can only produce a tidal heating luminosity of $10^{-5}-10^{-2}\,L_\odot$, much less than the nuclear luminosity of a helium-core-burning sdB (typically $\sim10\,L_\odot$). While on-resonance torques could be $2-3$ orders of magnitude larger, nonlinear dissipation likely makes the actual torque much weaker. Hence we do not expect tidal heating to ever surpass the stellar luminosity.

However, we note that tides do not dissipate energy uniformly inside the star. For gravity waves with radiative damping, the majority of tidal heating should occur near the stellar surface. Even a small amount of heat may change the stellar structure near the surface, which could in turn affect the dissipation processes in that region. Future works should investigate whether this scenario is important for the tidal evolution of sdB binaries.

\subsection{Other Caveats}
\label{sec:diss_caveats}

There are some other caveats with our methods. We calculate the oscillation modes based on pre-calculated nonrotating stellar models. As the star becomes spun up, the modes may start to behave differently. This matters the most when $\Omega_\mathrm{spin}$ becomes comparable with the Brunt V\"ais\"al\"a frequency $N_\mathrm{c}$ near the convective core boundary, where the oscillations are couple with the tidal potential. Rotation is also important when $\Omega_\mathrm{spin}$ becomes comparable with $\omega_\alpha$, where rotation effects on mode eigenfunctions cannot be ignored.

We checked that the former case is never important for sdB binaries, as $N_\mathrm{c}$ is typically greater than $10\,\mathrm{mHz}$, much higher than the rotation rates of even synchronized sdBs. For the latter case, as the mode contributing mostly to the tidal torque is the one with $\omega_\alpha\approx\omega_\mathrm{f}\equiv m(\Omega_\mathrm{orb}-\Omega_\mathrm{spin})$, the rotational effects on modes could only be significant when $\omega_\alpha\sim\Omega_\mathrm{orb}\sim\Omega_\mathrm{spin}$, i.e., when the system is close to synchronization. However, the tidal torques become small as synchronization is approached, so we do not expect these rotational effects to change our results much.

We ignored the tidal dissipation in the companion star in our spin-orbit evolution. This is valid for WD companions whose dissipation is negligible compared to sdB stars. However, for main sequence companions, tidal dissipation in the companion can occur. Since the orbital moment of inertia of the binary is always much greater than the spin moment of inertia of the companion, this process synchronizes the companion star without changing the orbital period significantly. The dissipation within the companion can thus be ignored if the mass and angular momentum of the binary are conserved.

An exception occurs if the companion star is undergoing magnetic braking. This may remove a significant amount of orbital angular momentum throughout the evolution. The coupled evolution of tides and magnetic braking have only been computed recently in some cases (see, e.g., \citealt{Sun2024}). Future works should investigate whether this is an important process with more detailed modeling of evolution including magnetic braking.

We set the initial orbital periods of our binary systems as a unified $P_\mathrm{orb}=60\,\mathrm{days}$, while in reality close-in sdB binaries can be born from two separate channels, namely a prior mass-transfer phase, or a common-envelope ejection. In general, sdB binaries born from these channels could have different initial spin periods, but the detailed outcome of these processes is highly uncertain. Nevertheless, we comment that, as long as tides are responsible for most of the sdB angular momentum for synchronized systems, our calculations should be insensitive to the initial orbital setup.

We did not include mass-transfer in the sdB evolution phase, even though it could happen for sdB binaries born at orbital periods less than $2-3$ hours \citep{Bauer2021}. Mass-transfer may remove the sdB outer hydrogen envelope, and hence change the strength of the tides (see discussion in Section \ref{sec:env_mass}). However, we suspect that tidal synchronization will remain efficient for these stars since they fill their Roche lobes.

\section{Conclusion}
\label{sec:conclusion}

In this manuscript, we investigated the tidal spin-up of close-in subdwarf B (sdB) binaries. We considered the dissipation of tidally excited gravity waves in the envelopes of sdB stars, and calculated the tidal torques by directly computing the amplitudes of tidally driven oscillation modes in sdB stellar models. We integrated the coupled spin-orbit evolution of these binaries and calculated the resulting sdB rotation rates.

We showed that in contrast to the usual assumption that gravity waves are efficiently damped near the surface (``Zahn's traveling wave limit''), these waves can actually be less damped, and can reflect back to form standing waves in the radiative envelope of sdB stars. The resulting tidal torque is then significantly less than Zahn's theory predicted, and has a complicated resonant dependence on the frequency of the tidal force. At longer periods, the waves are more highly damped and the tidal torque approaches Zahn's limit.

For binaries containing a $0.47\,M_\odot$ canonical sdB, our models predict the system will be tidally synchronized if the orbit is less than $\sim \! 0.2$ days. For those with a $0.37\,M_\odot$ low-mass sdB, this tidal synchronization period becomes $\sim \! 0.15$ days. These values are very similar to the observed spin rates of sdB binaries (Figure \ref{fig:obs_compare}), which are tidally synchronized at orbital periods less than $\sim \!0.2$ days. The tidal synchronization timescale has weak dependence on the companion star mass, and is mostly determined by the orbital period.

We investigated how the amount of hydrogen in the sdB envelope could affect the strength of the tidal torque. Since sdBs with more hydrogen have larger radii, and the tidal torque magnitude could scale with the stellar radius as $\tau\propto R^6$, tidal torques may be stronger for stars with more hydrogen. However, the existence of unstable oscillations for our sdB models with thicker hydrogen envelopes complicate the calculation of tidal torques.

When tidally synchronized sdBs evolve into carbon--oxygen white dwarfs, we estimate their rotation periods to be between 4 to 17 minutes (if tidal effects after the core-helium-burning phase can be neglected), which corresponds to spin rates roughy a hundred times faster than typical white dwarfs. We pointed out that resonance locking cannot happen in the tidal spin-up phase of sdB binaries, and discussed the limitations of our mode decomposition method to calculate tidal torques. Differential rotation, rotational effects on oscillations, and tidal heating may also be important. Future works should investigate the above scenarios, and compare them to the growing numbers of rotation rate measurements for sdBs in close binaries.

The agreement between our models and measurements for sdB binaries is very encouraging for the prospect of reliable tidal synchronization predictions. In particular, we expect the physics of tidal spin-up in sdBs to be very similar to that of more massive helium stars in close binaries \citep{Ma2023}, which are progenitors of gravitational wave sources and exotic supernovae. We believe the results of this paper increase the credibility of predictions for black hole spins presented in that work.

\section*{Acknowledgements}

We thank Holly Preece and the anonymous referee for constructive comments on the manuscript. We thank Evan Bauer, Ylva G\"otberg, Rich Townsend, Emily Hu, Peter Scherbak, Reed Essick, and Yuri Levin for helpful discussions. This work is partially supported by NASA through grant 20-XRP20 2-0147. L.M. is thankful for the Max Planck Institute for Astrophysics and the Kavli Foundation, who supported the 2023 Kavli Summer Program where many of the above discussions took place.

\software{MESA \citep{Paxton2011,Paxton2013,Paxton2015,Paxton2018,Paxton2019}, GYRE \citep{townsend:13,townsend2018,goldstein2020}}, SciPy \citep{scipy}, matplotlib \citep{Hunter2007}, numpy \citep{Harris2020}

\appendix
\section{Fixing MESA Profiles}
\label{app:fix_rho}

\begin{figure*}
    \centering
    \includegraphics[width=\textwidth]{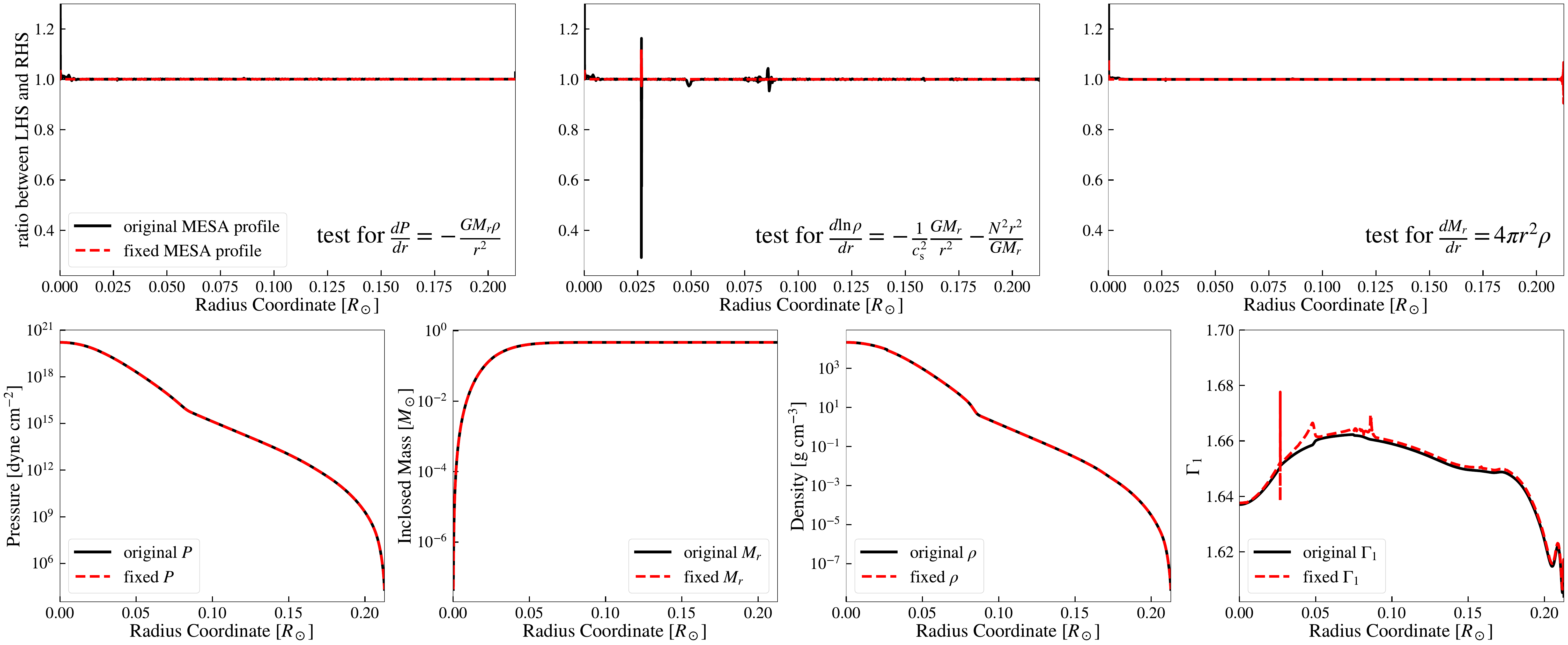}
    \caption{\textbf{Upper}: The ratios between the LHS and RHS of Equations \ref{eq:dP_dr}, \ref{eq:dM_dr} and \ref{eq:dlnrho_dr} for an example MESA snapshot, which should all be unity. We see that in the original stellar profile calculated from MESA, Equation \ref{eq:dlnrho_dr} is sometimes not satisfied, with the largest discrepancy happening at the convective core boundary with a density discontinuity. We fix the $P,M_r$ and $\rho$ profiles by the method described in Appendix \ref{app:fix_rho}, and the fixed profiles satisfy Equation \ref{eq:dlnrho_dr} better. \textbf{Lower}: The original and fixed $P,M_r, \rho$ and $\Gamma_1$ profiles of the example MESA snapshot. The pressure, density and enclosed mass are almost identical to original MESA profiles, so the change of stellar structure after the fixing process is negligible. $\Gamma_1$ is different by a small amount in some regions of the star, which means that the energy processes in the fixed stellar model are in general not consistent.}
    \label{fig:fix_rho}
\end{figure*}

To calculate the tidal response, we solve for internal stellar oscillations with the GYRE stellar oscillation code \citep{townsend:13,townsend2018,goldstein2020}. The code reads stellar snapshots from MESA as unperturbed background profiles of density, pressure, etc., and then solves the linear perturbation equations of stellar oscillations. The oscillation equations GYRE aims to solve are simplified by assuming hydrostatic equilibrium and mass conservation of the background stellar profile, hence the MESA snapshots provided to GYRE should satisfy the following equations:
\begin{align}
    \label{eq:dP_dr}
    \frac{dP}{dr}&=-\rho g\,,\\
    \label{eq:dM_dr}
    \frac{dM_r}{dr}&=4\pi r^2\rho\,,
\end{align}
where $P,\rho,M_r$ and $g=GM_r/r^2$ are the pressure, density, enclosed mass and gravity inside the star. Further, the Brunt-V\"ais\"al\"a frequency profile inside the stellar model should also satisfy the following equation by definition:
\beq
N^2\equiv g\bigg(\frac{1}{\Gamma_1}\frac{d\ln P}{dr}-\frac{d\ln\rho}{dr}\bigg)=g\bigg(-\frac{g}{c_\mathrm{s}^2}-\frac{d\ln \rho}{dr}\bigg)\,,
\eeq
where we made use of $c_\mathrm{s}^2\equiv\Gamma_1P/\rho$. We rewrite the above equation into the following form:
\beq
\label{eq:dlnrho_dr}
\frac{d\ln \rho}{dr}=-\frac{N^2}{g}-\frac{g}{c_\mathrm{s}^2}\,,
\eeq
and plot the ratios between the LHS and RHS of Equations \ref{eq:dP_dr}, \ref{eq:dM_dr} and \ref{eq:dlnrho_dr} for one of our MESA stellar snapshots in the upper panels of Figure \ref{fig:fix_rho}. While they should all be unity, we notice that in the original MESA profile, the ratio between the LHS and RHS for equation \ref{eq:dlnrho_dr} departs significantly from 1 at the density discontinuity near the convective core boundary at $0.025\,R_\odot$. In the radiative envelope, this ratio also departs from unity by a few percent at some radii. The GYRE oscillation solutions solved by assuming Equation \ref{eq:dlnrho_dr} are hence problematic. In practice, we find that this inconsistency often causes the mode solutions to change drastically between successive MESA snapshots, while in principle we expect them to vary gradually as the modes evolve.

We hence need to fix the stellar profiles provided by MESA to get correct oscillation solutions. As g mode oscillations are most sensitive to the Brunt-V\"ais\"al\"a frequency profile, we aim to keep the value of $N$ as output by MESA, and adjust the density and pressure profiles to satisfy Equations \ref{eq:dP_dr}, \ref{eq:dM_dr} and \ref{eq:dlnrho_dr}. Hence, we rewrite these equations into the following matrix form:
\beq
\label{eq:matrix}
\frac{d}{dr}
\begin{bmatrix}
P\\
\ln\rho \\
M_r
\end{bmatrix}
=
\begin{bmatrix}
-\frac{GM_r\rho}{r^2}\\
-\frac{N^2r^2}{GM_r}-\frac{GM_r}{r^2c_\mathrm{s}^2}\\
4\pi r^2\rho
\end{bmatrix}\,.
\eeq
The equations then become a first order ordinary differential equation of the form $d\mathbf{y}/dr=f(\mathbf{y},r;N^2,c_\mathrm{s}^2)$ where $\mathbf{y}(r) = [P(r),\ln\rho(r),M_r(r)]$ is an unknown function to be solved numerically, and $N^2$ and $c_\mathrm{s}^2$ are the ODE parameters that can be fitted from the original stellar profile. To solve for $\mathbf{y}$, we need a set of boundary conditions for $P,\rho$ and $M_r$, which are given by the following physical limits:

\begin{align}
\label{eq:bdrP}
    &P(R_\mathrm{star})=P_\mathrm{original}(R_\mathrm{star})\,;\\
    \label{eq:bdrrho}
    &\rho(\epsilon)=\rho_\mathrm{original}(0)\,;\\
    \label{eq:bdrm}
    &M_r(\epsilon)=\frac{4\pi\rho_\mathrm{original}(0)}{3}\epsilon^3\,,
\end{align} 
Here, $\epsilon$ is the radial coordinate of a point very near the center of the star, just above $r=0$ to avoid the singularity at $r=0$ which can cause numerical problems. Once the solutions are found, the values at $r=0$ are acquired by $P(0)=P(\epsilon)$, $\rho(0)=\rho(\epsilon)$ and $M_r(0)=0$. We choose $\epsilon$ to be the spatial coordinate of the innermost grid in the MESA model, and we can then solve Equation \ref{eq:matrix} with the boundary conditions described by Equations \ref{eq:bdrP}, \ref{eq:bdrrho} and \ref{eq:bdrm}. 

We use the \texttt{integrate.solve\_bvp} function in the SciPy python package \citep{scipy} to solve for the fixed $P,M_r$ and $\rho$ profiles, using the original MESA profiles as our initial guess. We show the comparison between our fixed and original MESA profiles in Figure \ref{fig:fix_rho}.
The fixed $P,M_r$ and $\rho$ profiles are almost identical to the original density profile, but they more accurately satisfy Equation \ref{eq:dlnrho_dr} in the radiative envelope. While there is still some inconsistency at the convective core boundary due to the density discontinuity, it is likely unimportant as gravity waves are evanescent in the convective core.

We note that, as the adiabatic index $\Gamma_1\equiv\rho c_\mathrm{s}^2/P$ is now calculated from the fixed $P$ and $\rho$ and the original sound speed profile, it can be different from the original $\Gamma_1$ by a small amount inside the star. Physically, this means the gas equation of state is now slightly artificial, and future works should investigate a more self-consistent way to deal with this problem. Nevertheless, we find that, after fixing the MESA profiles, GYRE is able to get oscillation solutions that vary gradually across nearby MESA snapshots.

\bibliography{sdB}
\bibliographystyle{aasjournal}

\end{CJK*}
\end{document}